\documentclass[reprint,amsmath,amssymb,superscriptaddress,aps,pre]{revtex4-2}
\usepackage{geometry}
\usepackage{comment}
\usepackage{enumerate}
\usepackage{amsmath,nccmath}
\usepackage{xcolor}
\usepackage{mathrsfs}
\usepackage{graphicx}
\usepackage{dcolumn}

\usepackage{natbib}
\usepackage{bm}

\usepackage[mathlines]{lineno}
\usepackage{xcolor}
\usepackage{color}
\usepackage[colorlinks=true,linktoc=page,citecolor=black,linkcolor=black,]{hyperref}
\newcommand{\bb}{{\bm b}}
\newcommand{\bu}{{\bm u}}
\newcommand{\bx}{{\bm x}}
\newcommand{\bp}{{\bm p}}
\newcommand{\bq}{{\bm q}}
\newcommand{\br}{{\bm r}}
\newcommand{\bj}{{\bm j}}

\newcommand{\ba}{{\bm a}}

\newcommand{\bmm}{{\bm m}}
\newcommand{\bomega}{\boldsymbol \omega}

\newcommand{\lamb}{{\bm u\times\bomega}}
\newcommand{\lambp}{{\bm u'\times\bomega'}}
\newcommand{\jb}{{\bm j\times\bm b}}
\newcommand{\jbp}{{\bm j'\times\bm b'}}
\newcommand{\ub}{{\bm u\times\bm b}}
\newcommand{\ubp}{{\bm u'\times\bm b'}}
\begin{document}

\title{Universal turbulent relaxation of fluids and plasmas by the  principle of vanishing nonlinear transfers} 
\author{Supratik Banerjee}
\email{sbanerjee@iitk.ac.in}
\author{Arijit Halder}
\author{Nandita Pan}
\affiliation{
 Department of Physics, Indian Institute of Technology Kanpur, Uttar Pradesh, 208016 India
}%
\date{\today}
\begin{abstract}
A seventy year old problem of fluid and plasma relaxation has been revisited. A new principle of vanishing nonlinear transfer has been proposed to develop a unified theory of turbulent relaxation of neutral fluids and plasmas. Unlike previous studies, the new principle enables us to find the relaxed states unambiguously without going through any variational principle. The general relaxed states obtained herein are found to support naturally a pressure gradient which is consistent with several numerical studies. 
Relaxed states are reduced to Beltrami aligned states where the pressure gradient is neglected. \textcolor{black}{According to the present theory, the relaxed states are attained in order to maximize a fluid entropy $\mathcal{S}$ calculated from the principles of statistical mechanics (Carnevale \textit{et.al.}, J. Phys. A: Math. Theor., 1981).} This method can be extended to find the relaxed states for more complex flows. 
\end{abstract}
\maketitle

Self-organizing dynamic relaxation in neutral fluids and plasmas is an old but scarcely understood subject. Although a considerable number of works have already been accomplished to explain the relaxed states in different flows, an unambiguous definition of such a state and a universal physical principle to achieve the same has yet to be agreed upon. Despite this fact, a relaxed state is often analytically obtained by extremizing (minimizing or maximizing) a target function (TF) subject to one or more constraints of the flow. Initially the observed alignment (also called Beltrami-Taylor state, BT state hereinafter) between the magnetic field \footnote{expressed in Alfv\'en units} $\bb$ and the current field $\bj\ ( =\bm{\nabla}\times\bb)$ in cosmic plasmas {\it i.e.} $\bj = \lambda \bb$, (where $\lambda$ is a scalar function of space) was analytically obtained by maximizing the total magnetic energy for a given mean square current density \cite{Chandrasekhar1958}. Later a similar state was obtained in a more convincing way by minimizing the magnetic energy for a constant magnetic helicity and $\lambda$ was shown to be a global constant of the system \cite{Woltjer1958a, Taylor1974}. One popular way to find the aligned states is based on the principle of selective decay where the relaxed states are obtained by varying the rapidly decaying quantity (chosen as the TF) subject to the invariance of the slowly decaying quantities (chosen as the constraints). For 3D incompressible magnetohydrodynamics (MHD), the rate of decay of the total energy $E \ (= \int \frac{1}{2} (u^2 + b^2) \ d\tau $) is found to be greater than that of the two helical invariants namely the cross helicity $H_C \ (=\int \bu\cdot\bb \ d \tau$) and the magnetic helicity $H_M \ (=\int \ba\cdot\bb \ d \tau$), where $\bu$ and $\ba$ represent the fluid velocity and the magnetic vector potential respectively and the integration is done over the space. The self-organized states can therefore be obtained by varying  
\begin{equation}
E - \lambda_1 H_M - \lambda_2 H_C \label{eq3b2}
\end{equation}
{\it w.r.t.} $\ba$ and $\bu$ respectively, where $\lambda_{1,2}$ denote the undetermined multipliers of Lagrange. Such a variation finally gives the relaxed configurations as
\begin{align}
 \bm{\nabla}\times\bb &= 2 \lambda_1\bb+\lambda_2\bomega \,\,\, \text{and}\label{eq3b4}\\
 \bu &= \lambda_2\bb, \label{eq3b5}
\end{align}
where $\bomega = \bm{\nabla} \times \bu$.
Solving the above equations \eqref{eq3b4} and \eqref{eq3b5}, we get $\lambda_1=0$ and $\lambda_2=\pm 1$ which, in turn, exactly correspond to the states
\begin{equation}
\bu = \pm \bb, \quad \text{and hence} \quad
\bj\times\bb = \bomega\times\bu, \label{eq3b6}
\end{equation}
previously obtained in \cite{Woltjer1958b}. Note that, for incompressible MHD, a false BT alignment condition may seem to be obtained if one substitutes 
Eq.~\eqref{eq3b5} in Eq.~\eqref{eq3b4} without solving for $\lambda_{1,2}$. Such a possibility is obviously eliminated as $\lambda_1$ vanishes. This clearly suggests that an alignment between $\bb$ and $\bj$ is only possible when $\bu$ and $\bomega$ are aligned. Using similar formalism, relaxed states were also obtained for 3D Hall MHD (HMHD), where apart from $E$ and $H_M$, the total generalised helicity $H_G \ (= \int (\ba + d_i \bu)\cdot(\bb + d_i \bomega)\ d\tau)$ is also an inviscid invariant ($d_i$ being the ion inertial length). The relaxed states can be obtained by varying
\begin{equation}
    E - \lambda_1 H_M - \lambda_2 H_G \label{hmhdvar}
\end{equation}
{\it w.r.t.} $\ba$ and $\bu$ thereby leading to 
\begin{align}
\bm{\nabla}\times\bb &= 2\left(\lambda_1+\lambda_2\right)\bb + 2\lambda_2 d_i \bomega\,\,\,\text{and}\label{eq3hb3} \\
\bu &= 2\lambda_2 d_i \bb + 2\lambda_2 d_i^2 \bomega \label{eq3hb4}
\end{align}
respectively. Further simplification leads to well-known double-curl Beltrami states for HMHD plasmas given by \cite{Mahajan1998, Yoshida1999}
\begin{equation}
\bm{\nabla}\times\left(\bm{\nabla}\times\bb\right) - \alpha\left(\bm{\nabla}\times\bb\right) + \beta \bb = 0, \label{eq3hb6}
\end{equation}
with $\alpha = (1+4\lambda_1 \lambda_2 d_i^2)/2\lambda_2 d_i^2$ and $\beta = (\lambda_1 + \lambda_2)/\lambda_2 d_i^2$. However, the variational problem in Eq.~\eqref{hmhdvar} is mathematically ill-posed as the decay rate of $H_G$ may supersede that of $E$, and a mere permutation of $E$ and $H_G$ cannot solve this problem \cite{Yoshida2002}. To get rid of this issue, the generalized enstrophy was chosen as the desired TF and through its variation a triple-curl Beltrami state in $\bb$ was obtained. In contrast to the ordinary MHD, one can immediately see that the above relaxed state permits BT alignment as a natural solution. Without using Taylor's selective decay hypothesis, an interesting theory of BT relaxation was also proposed for resistive MHD using Cauchy-Schwartz inequality \cite{Qin2012}.

Despite the previous works, it has been observed that the relaxed state of an MHD plasma is rather given by a force-balanced minimum energy state supporting a finite pressure gradient as \cite{Zhu1995, Zhu1996, Sato1996}
\begin{equation}
\jb = \bm{\nabla}p.
\end{equation}
Such a state can trivially be obtained as the solution of a hydrostatic equilibrium. However, to explain this in general, a complimentary approach was implemented by using the principle of minimum entropy production rate (MEPR) \cite{Prigogine1955}. While for a low-$\beta$ plasma, the BT state was approximately recovered using MEPR, a relaxed hydrodynamic state supporting finite pressure gradient was analytically obtained later using the same principle \cite{Ting1986, Hamieri1987, Montgomery1988, Montgomery1989}. 
In particular, using complex Chandrasekhar-Kendall functions, such a state was also justified from a triple-curl Beltrami alignment
\begin{equation}
\bm{\nabla}\times\bm{\nabla}\times\left(\bm{\nabla}\times\bb\right) = \lambda \bb
\end{equation}
in the absence of the mean plasma flow \cite{Dasgupta1998}. Although the principle of MEPR appears to be less ambiguous and more general than the method of selective decay, it is only able to describe the evolution of the states close to the states of relaxation. In addition, the previous works only considered low-$\beta$ plasmas, and hence a complete description of a plasma relaxation is still lacking \cite{Hamieri1987, Montgomery1988}. 

Finding the relaxed states for a 3D hydrodynamic (HD) flow is tricky. Such a system permits two inviscid invariants namely, the total kinetic energy $E_K\ (=\int \frac{1}{2}u^2 d\tau)$ and the total kinetic helicity $H_K \ (=\int \bu\cdot\bomega d\tau)$. A Beltrami-type aligned state $\bu = \lambda \bomega$ can simply be obtained by varying $E_K$ for a constant $H_K$. However as discussed previously, such a variation is mathematically ill-posed as $H_K$ may have a higher decay rate than $E_K$. One then needs a TF that decays faster than both $E_K$ and $H_K$. The total enstrophy, $\Omega \ (=\int \omega^2 \ d \tau)$ indeed serves this purpose and varying this with $E_K$ and $H_K$ as constraints, we obtain 
\begin{equation}
    \bm{\nabla} \times \bomega = \frac{{\lambda_1}}{2} \bu + \lambda_2 \bomega, \label{hd1}
\end{equation}
which evidently permits $\bu$-$\bomega$ alignment as a possible solution. Interestingly, for the HD case the above variational principle and the subsequent relaxed states in Eq.~\eqref{hd1} can also be obtained using MEPR. Note that, a $\bu$-$\bomega$ aligned state was also obtained by varying $\Omega$ while keeping $H_K$ as the only constraint \cite{Gonzalez2008, Gonzalez2010}. However, their work used a heuristic $\bb$-$\bomega$ analogy in the variational principle originally proposed by \cite{Taylor1974} and was inconclusive about the meaning of such relaxation. Similar to the 3D MHD case, the relaxed state of a 3D HD flow is also found to relax towards a state with a finite pressure gradient as \cite{Kraichnan1988, She1991}
\begin{equation}
\lamb = \bm{\nabla} p,
\end{equation}
and unfortunately such a state has not been theoretically obtained till date. As summarized, various competing theories of plasma relaxation have been being proposed for the last seventy years. Most of them predicted aligned relaxed states using variational principle pivoted on different perspectives thus leading to a non-unique choice of the TF and the constraints. 
Interestingly, a considerable drop of the nonlinear terms in the evolution equations was observed numerically \cite{Matthaeus1980, Ting1986, Stribling1991, Kraichnan1988, Servidio2008, Tsinober1999} and it was realized that a dynamic relaxed state should be 
`as free from turbulence as possible' \cite{Yoshida2002}. Nevertheless, a universal theory of turbulent relaxation in fluids and plasmas has not been developed to the best of our knowledge. 

\begin{figure*}
    \centering
    \includegraphics[width=16cm, height= 6cm]{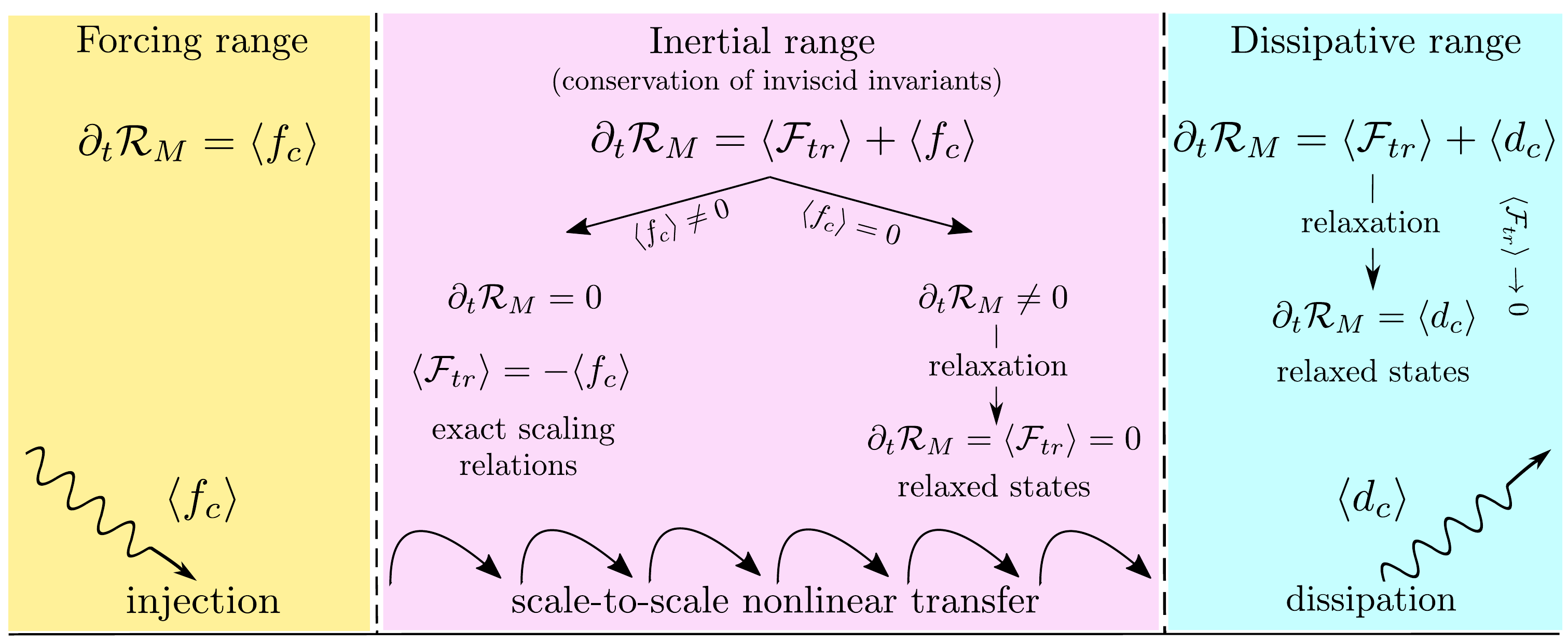}
    \caption{Schematic diagram for principle of vanishing nonlinear transfer}
    \label{schematic}
\end{figure*}

In this Letter, we concentrate on the dynamic relaxed states of a turbulent flow and propose a universal way to characterize such states in both neutral fluids and plasmas. By definition, turbulence is an out of equilibrium flow regime dominated by nonlinearity where the system conceives a large number of length and time scales. If $M = \int \left({\bp \cdot \bq}\right) d \tau $ is an inviscid invariant of the flow, then $\partial_t \langle\bp\cdot\bq\rangle = \langle{\mathcal{F}}_{M}\rangle + \langle d_{M}\rangle + \langle f_{M}\rangle$,
where $\mathcal{F}_{M}$ is the flux term, $d_{M}$ the dissipative term, $f_{M}$ the forcing term and $\langle\cdot\rangle$ denotes the statistical average which becomes identical to the space average for homogeneous turbulence. ${\mathcal{F}}_{M}$ can be written as a pure divergence term which vanishes due to Gauss divergence theorem, leading to a statistical stationary state given by $\langle d_{M}\rangle = -\langle f_{M}\rangle$. For scale-dependent transfers, one has to consider the evolution of ${\cal R}_M = \left\langle \frac{\bp \cdot \bq' + \bp' \cdot \bq }{2} \right\rangle$, which is the symmetric two-point correlator of $M$. Here, the unprimed and primed quantities represent the corresponding field properties at point $\bx$ and $\bx'(\equiv\bx + \br)$ respectively and are independent of each other. For homogeneous turbulence, any correlation function of primed and unprimed variables becomes scale-dependent \textit{i.e.} a function of $\br$ only. The evolution equation of the correlator ${\cal R}_M$ can be written as 
\begin{equation}
    \partial_t{\cal R}_M = \langle\mathcal{F}_{tr}^M\rangle + \langle f_c^M\rangle + \langle d_c^M\rangle, \label{tpflux}
\end{equation}
where $\langle\mathcal{F}_{tr}^M\rangle$, $\langle f_c^M\rangle$ and $\langle d_c^M\rangle$ represent the scale-dependent rates of nonlinear transfer, injection and dissipation of $M$ respectively. 
Near the injection scale, Eq.~\eqref{tpflux} reduces to $\partial_t{\cal R}_M = \langle f_c^M\rangle$ and a stationary state can be achieved when $\langle f_c^M \rangle = 0$ (decaying turbulence). For the so-called inertial range, where $\langle f_c^M \rangle$ is taken as a constant input (like a uniform background) and $\langle d_c^M \rangle$ can be neglected, Eq.~\eqref{tpflux} reduces to $\partial_t{\cal R}_M = \langle\mathcal{F}_{tr}^M\rangle + \langle f_c^M\rangle$.
Now for $\langle f_c^M\rangle \neq 0$, a stationary state leads to $\langle\mathcal{F}_{tr}^M\rangle = -\langle f_c^M\rangle$ and we obtain the exact relations in forced stationary turbulence \cite{Monin1975, Politano1998b, Banerjee2016a, Banerjee2016b, Mouraya2019, Pan2022}. If the energy input is removed, {\it i.e.} $\langle f_c^M\rangle = 0$, \textcolor{black} {then for all scales inside the inertial range, a non-stationary transient state is achieved as $\partial_t{\cal R}_M = \langle\mathcal{F}_{tr}^M\rangle\neq 0$. Since, the inertial range length scales can neither inject nor dissipate but can only nonlinearly transfer invariants to the subsequent scales, it is reasonable to expect that a trivial steady state is achieved at relaxation where $\langle\mathcal{F}_{tr}^M\rangle$ vanishes.}
Such a state is called a `relaxed state' in the premise of our proposed principle which we call the principle of vanishing nonlinear transfer (PVNLT hereinafter). For dissipative scales,  $\langle\mathcal{F}_{tr}^M\rangle$ can be considered as the input and hence one can write $\partial_t{\cal R}_M = \langle\mathcal{F}_{tr}^M\rangle + \langle d_c^M\rangle$. A relaxed state $(\langle\mathcal{F}_{tr}^M\rangle = 0)$ therefore implies a non-stationary dissipative state for small scales. According to PVNLT, a turbulent system attains a non-static relaxed state in order to maintain the statistical stationarity \textcolor{black}{for two-point correlators} at all scales within the inertial range. 
A schematic diagram of the discussed principle is given in FIG.\eqref{schematic}.

\textcolor{black}{ The abovesaid macroscopic principle can indeed be explained using the principles of statistical mechanics. One can indeed formulate a Boltzmann H-theorem for ideal incompressible fluids and plasmas having a spectral cutoff \footnote{The Imposition of a cutoff is consistent with the dissipative anomaly of turbulence theory.}\citep{Carnevale1981}. Such systems always try to maximize a fluid entropy functional $\mathcal{S}$. For a turbulent system with inviscid invariant $M$, $\mathcal{S} \equiv \mathcal{S}[\widehat{\mathcal{R}}_M(\bm{k})]$, where $\widehat{\mathcal{R}}_M(\bm{k})$ is the Fourier transform of $\mathcal{R}_M$. 
Using second-order Markovian closure, it is shown that $d\mathcal{S}/dt \geq 0 $. According to our theory, a relaxed state is obtained when $\mathcal{S}$ attains its maximum value. For a relaxed state, inside the inertial range, we therefore have $d\mathcal{S}/dt = 0 \implies \partial_t {\cal \widehat{R}}_M(\bm{k}) = 0\implies \partial_t \mathcal{R}_M= 0\implies  {\langle\mathcal{F}^M_{tr}\rangle = 0}$, thereby entailing PVNLT. }
As we shall see, this definition would help us in obtaining the aforementioned relaxed configurations in both neutral fluids and plasmas in a systematic manner. For 3D MHD flow, we define the symmetric two-point correlators for $E$, $H_M$ and $H_C$ as ${\cal R}_E = \langle \bu \cdot \bu' + \bb \cdot \bb' \rangle/2$, ${\cal R}_{H_{M}} = \langle \ba \cdot \bb' + \ba' \cdot \bb\rangle/2$ and ${\cal R}_{H_C} = \langle \bu \cdot \bb' + \bu' \cdot \bb \rangle/2$, respectively and the corresponding evolution equations are written as 
\begin{align}
\partial_t{\cal R}_E &= \langle\mathcal{F}_{tr}^E\rangle + \langle f_c^E\rangle + \langle d_c^E\rangle,\\
\partial_t{\cal R}_{H_M} &= \langle\mathcal{F}_{tr}^{H_M}\rangle + \langle f_c^{H_M}\rangle + \langle d_c^{H_M}\rangle, \\
\partial_t{\cal R}_{H_C} &= \langle\mathcal{F}_{tr}^{H_C}\rangle + \langle f_c^{H_C}\rangle + \langle d_c^{H_C}\rangle,
\end{align}
where,
{\small
\begin{widetext}
\begin{align}
\langle\mathcal{F}_{tr}^E\rangle &= \frac{1}{2}\left\langle \bu'\cdot(\lamb+\jb - \bm{\nabla} P_{ T})+ \bu \cdot(\lambp+\jbp - \bm{\nabla}' P_T')+ \bb'\cdot\bm{\nabla}\times(\ub) + \bb\cdot\bm{\nabla}'\times (\ubp)\right\rangle,\label{tpfluxE} \\
\langle\mathcal{F}_{tr}^{H_M}\rangle &= \frac{1}{2}\left\langle \ba'\cdot\bm{\nabla}\times(\ub)+ \ba\cdot\bm{\nabla}'\times(\ubp) + \bb'\cdot(\ub) + \bb\cdot(\ubp) \right\rangle, \label{tpfluxHm}\\
\langle\mathcal{F}_{tr}^{H_C}\rangle &= \frac{1}{2}\left\langle \bb'\cdot(\lamb+\jb - \bm{\nabla} P_T)+ \bb \cdot (\lambp+\jbp - \bm{\nabla}' P_T')  
+ \bu'\cdot\bm{\bm{\nabla}}\times (\ub) + \bu\cdot\bm{\bm{\nabla}}'\times (\ubp) \right\rangle,\label{tpfluxHc}
\end{align}
\end{widetext}
}
\noindent with $P_T = p + u^2/2$ and omitting the gauge term in $\partial_t \ba$. As per our definition above, for a relaxed state we have $\langle\mathcal{F}_{tr}^E\rangle=\langle\mathcal{F}_{tr}^{H_M}\rangle=\langle\mathcal{F}_{tr}^{H_C}\rangle=0$ at all scales within the inertial range. Furthermore, in homogeneous turbulence, for any solenoidal vector field $\bmm$ and scalar function $\theta$, we have  $\left\langle\bmm'\cdot\left(\bm{\nabla}\theta\right)\right\rangle = -\left\langle\theta(\bm{\nabla}'\cdot\bmm')\right\rangle=0$. Using the aforementioned facts, for a non-trivial relaxed state (where none of the invariant vanishes identically), one should simultaneously have
\begin{align}
\lamb+\jb &= {\bm{\nabla}} \left(P_T + \phi_0\right)\label{3dmhdalign1} \,\,\,\,\,\, \text{and}\\
\ub &= \bm{\nabla} \psi_0,\label{3dmhdalign2}
\end{align}
where $\phi_0$ and $\psi_0$ are arbitrary scalar fields. The determination of $\phi_0$ and $\psi_0$ is system-specific. An alignment between $\bu$ and $\bb$ is usually observed in space plasmas {\it e.g.} solar wind \cite{Riley1995, Wicks2013}, leading to the choice $\bm{\nabla}\psi_0 = {\bf 0}$, which gives
\begin{align}
\lamb+\jb &= {\bm{\nabla}} \left(P_T + \phi_0\right) \,\,\,\,\,\, \label{alignuwjb}\text{and}\\
\ub &= \bf{0}. \label{alignub}
\end{align}
From Eq.~\eqref{alignub}, we have $\bu = \lambda \bb$ and using the fact that $\lambda$ is a global constant, we have $\bomega = \lambda \bj$ and 
\begin{equation}
\jb = \frac{\bm{\nabla} \left(P_T+ \phi_0\right)}{1- \lambda^2}. 
\end{equation}

\textcolor{black}{Furthermore, neglecting $\bm{\nabla}\phi_0$ and using the identity $\bj\times\bb=\left(\bb\cdot\bm{\nabla}\right)\bb-\bm{\nabla}\left(b^2/2\right)$, the above equation can be further reduced to
\begin{equation} \left(\bb\cdot\bm{\nabla}\right)\bb=\frac{\bm{\nabla}\left(p+u^2/2+(1-\lambda^2)b^2/2\right)}{1-\lambda^2}.\label{BTred}
\end{equation}
For an incompressible low-${\beta}$ plasma (${p\ll |\bm{b}|^2/2}$) with negligible flow inertia (${|\bm{u}|\ll |\bm{b}|}$) \textit{i.e.} ${1-\lambda^2\approx 1}$), the above equation reduces to ${(\bm{b}\cdot\bm{\nabla})\bm{b}\approx\bm{\nabla}(b^2/2)}$, thus resulting in a BT aligned state where $ {\jb \approx \bm{0}}$.} Note that, in previous studies where higher order multi-curl Beltrami states were obtained as a result of an extremization principle \cite{Mahajan1998, Dasgupta1998, Yoshida1999, Yoshida2002, Bhattacharyya2003}, the aligned states could not be obtained as a natural limit of a relaxed state supporting the pressure gradient. However, in the current case, the relaxed states with pressure gradient emerge naturally and reduce to an aligned state in the appropriate limit. For the case of Alfv\'enic alignment $(\lambda = \pm 1)$, one obtains $\bm{\nabla} P_T ={\bf 0}$, thereby leading to $\lamb+\jb = {\bf 0}$. In presence of the Hall term, similar as above, one can also construct two-point correlator ${\cal R}_{E}$, ${\cal R}_{H_M}$ and ${\cal R}_{H_G}$ corresponding to the inviscid invariants.
For a relaxed state,
\begin{align}
\lamb+\jb &= \bm{\nabla} (P_T+ \phi_1)\,\,\,\, \text{and}\\
\left(\bu - d_i\bj\right)\times\bb &= \bm{\nabla} \psi_1.
\end{align}
For low plasma-$\beta$ and assuming $\bm{\nabla} \phi_1 = \bm{\nabla} \psi_1 ={\bf 0}$, the above two equations lead to
\begin{align}
    \bu - d_i \bj &= \lambda_1 \bb \,\,\, \text{and}\\
    \bb + d_i \bomega &= \lambda_2 \bu,
\end{align}
which are identical to the states obtained in Eq.~(10) of \citep{Mahajan1998}. Further calculations leads to a double-curl Beltrami state similar to Eq.~\eqref{eq3hb6} and to Eq.~(11) of \citep{Mahajan1998}.
\textcolor{black}{ Interestingly, our proposed relaxation principle can be shown to be consistent with numerically observed states obtained under certain initial conditions. It is well known that for a strongly helical system the final state is force-free whereas, for high initial alignment the system ends up in a Alfvenic state \cite{Stribling1991}. The same results can be obtained through PVNLT as we explain below:}

\textcolor{black}{For a strongly helical system, $\ba$ and $\bb$ are highly aligned, thus one can take $|\ba\times\bb|\sim\bm{0}$ which implies $|\bj\times\bb|\sim\bm{0}$. Hence, we can drop all the terms containing $\jb$ from Eqs.~\eqref{tpfluxE}-\eqref{tpfluxHc}. The relaxed states are obtained as 
\begin{align}
    \lamb&=\bm{\nabla}(P_T+\phi_0)\,\,\,\text{and}\\
    \ub&=\bm{\nabla}\psi_0.
\end{align}
Assuming $\bm{\nabla}\phi_0=\bm{\nabla}\psi_0 = \bm{0}$ and combining above two equations we get $\jb =\bm{\nabla}P_T/\lambda^2$, where $\lambda(\neq 0)$ is a constant. In the limit of low plasma-$\beta$, the given state further reduces to a BT aligned state $\jb=\bm{0}$. Similarly, for large initial alignment of $\bu$ and $\bb$, $|\bu\times\bb|\sim\bm{0}$. The relaxed state obtained in this case is given by $\lamb+\jb=\bm{\nabla}(P_T+\phi_0)$. Again, assuming $\bm{\nabla}\phi_0 =\bm{0}$ and low plasma-$\beta$, the relaxed state reduces to $\lamb+\jb=\bm{0}$ which is the most general solution. Now, high alignment between $\bu$ and $\bb$ implies $\bu=\lambda\bb$ and the general state reduces to $(1-\lambda^2)\jb=\bm{0}$. If initially one chooses $H_M$ to be low enough then $\jb$ cannot be neglected and we get $\lambda=\pm 1$, leading to $\bu=\pm\bb$ (Alfvenic state).
}

For ordinary hydrodynamics, the correlators for $E_K$ and $H_K$ are written as ${\cal R}_{E_K} = \left\langle {\bu \cdot \bu'} \right\rangle/2$ and ${\cal R}_{H_K} = \left\langle {\bu \cdot \bomega' + \bu' \cdot \bomega } \right\rangle/2$, respectively. In the relaxed state, the vanishing nonlinear transfer leads to
\begin{equation}
    \lamb = \bm{\nabla} \left(P_T + \phi_2\right).\label{alignuw}
\end{equation}
Again one can assume $\bm{\nabla} \phi_2 = {\bf 0}$ and the above state reduces to, $\lamb = \bm{\nabla} P_T$. As mentioned previously, such a relaxed state has been observed numerically in \citep{Kraichnan1988, She1991}. Unlike the MHD case, here a Beltrami alignment between $\bu$ and $\bomega$ is not easily found.

In contrast to three-dimensional flows, the relaxed states in two-dimensions are occasionally investigated \cite{Hasegawa1985}. In case of 2D hydrodynamics, $\bomega$ is perpendicular to the plane of $\bu$ and therefore $H_K$ vanishes identically at every point. The enstrophy $\Omega$ is a new inviscid invariant along with $E_K$. A relaxed state was obtained through the variational principle by varying $\Omega - \lambda_1 E_K$,
{\it w.r.t.} $\bu$, thereby leading to a double-curl Beltrami state in $\bu$, given by
\begin{equation}
   \bm{\nabla}\times\bomega = \lambda_1 \bu.\label{align2d1}
\end{equation}
It is easy to see that the above state also supports a $\bu$-$\bomega$ alignment as a possible solution \footnote{Following \cite{Yoshida2002}, a richer class of relaxed states may be obtained if the palinstrophy $P\  (=\int \left(\nabla \times {\bomega} \right)^2 d \tau)$ is varied for a fixed value of $E_K$ and $\Omega$. Such a variation would lead to a quadruple-curl Beltrami state in $\bu$ given by $\nabla \times \nabla \times \left( \nabla \times \bomega \right)= {\lambda_1} {\bu} + {\lambda_2} \left(\nabla \times \bomega \right)$ which permits Eq.~\eqref{align2d1} as a possible solution.}. Similar to $H_K$, magnetic helicity $H_M$ also vanishes trivially in 2D MHD. Instead, mean square vector potential $A \ (= \int {a}^2 d \tau)$ is conserved along with $E$ and $H_C$. The relaxed states are obtained by varying $E - \lambda_1 A - \lambda_2 H_C$,
{\it w.r.t.} $\ba$ and $\bu$ respectively, thereby leading to 
\begin{align}
 \bm{\nabla}\times\bb &=  2\lambda_1\ba+\lambda_2\bomega\,\,\,\text{and}\label{align2dbw}\\
 \bu &= \lambda_2\bb. \label{align2dub}
\end{align}
Combining Eqs.~\eqref{align2dbw} and \eqref{align2dub} one obtains  
\begin{equation}
    \bm{\nabla} \times \left(\bm{\nabla} \times \ba \right) = \lambda \ba,\label{2dmhdalign}
\end{equation}
where $\lambda= 2 \lambda_1/\left(1- {\lambda_2}^2 \right)$.
Our proposed principle can be extended, without any problem, for two-dimensional flows as well. The symmetric two-point correlator for $\Omega$ is defined as ${\cal R}_{\Omega} = \left\langle \bomega \cdot \bomega'\right\rangle$. From $\partial_t {\cal R}_{E_K}$ and $\partial_t{\cal R}_{\Omega}$, at relaxed state, one obtains, $\bm{\nabla}\times\left(\lamb\right) = {\bf 0}$ (similar to 3D case). For a two-dimensional flow, further we have 
\begin{equation}
    \bm{\nabla}\times\left(\lamb\right)=-\left(\bu\cdot\bm{\nabla}\right)\bomega=\bu\times\left(\bm{\nabla}\times\bomega\right) = {\bf 0},
\end{equation}
leading to double-curl Beltrami state in $\bu$. This is in agreement with the relaxed state obtained due to minimization of $\Omega$ for a given $E_K$ \cite{Hasegawa1985}. 
The correlator for $A$ is written as ${\cal R}_{A} = \left\langle \ba\cdot\ba'\right\rangle$. 
From $\partial_t{\cal R}_{E}$, $\partial_t{\cal R}_{A}$ and $\partial_t{\cal R}_{H_C}$, for the relaxed state one obtains similar conditions as given in Eqs.~\eqref{alignuwjb} and \eqref{alignub}. For a 2D flow, $\ba$ is perpendicular to the plane of the flow containing $\bu$ and $\bb$. One can therefore say $\ba \times \left(\bm{\nabla}\times\bu\right) = {\bf 0}$. Since from Eq.~\eqref{alignub}, $\bu = \lambda \bb$, the relaxed condition is given by $\ba \times \left(\bm{\nabla}\times\bb\right) = {\bf 0}$, leading to a double-curl Beltrami state in $\ba$, similar to Eq.~\eqref{2dmhdalign}. Note that, the study of HMHD flow strictly in two dimensions leads to an inconsistency in the evolution equation of the vector potential $(\partial_t \ba)$. To get out of this issue, numerical studies have been done for 2.5D HMHD \cite{Donato2012, Wang2012, Papini2021} where, the velocity and the magnetic fields have three components without any functional dependence on $z$. The relaxed states for such a system are exactly similar to those obtained for a 3D HMHD flow.

The present work proposes a simple and fundamental solution to the long-standing problem of dynamic relaxation of fluids and plasmas in terms of PVNLT. \textcolor{black}{ The proposed principle is universal for incompressible fluids and plasmas consistent with a high Reynolds number turbulence regime. The BT aligned states are obtained in the limit of insignificant pressure gradient. Unlike the previous approaches, our theory does not use the principle of selective decay and explains the dynamic relaxation as a state of maximum fluid entropy functional $S$ and naturally connects the relaxed states with and without the pressure gradient.} Note that, for obtaining the relaxed states using PVNLT, one needs to have the prior knowledge of all the inviscid invariants. However, unlike the method of selective decay, here, we do not require to compare the decay rates of those quantities in the presence of dissipation. Furthermore, our methodology is not affected by the direction of the cascades. Unlike the principle of MEPR, our analysis is not depending on the perturbation of states close to equilibrium. Interestingly, the alternative form of exact relations in turbulence directly shows that the turbulent flux vanishes in the relaxed states obtained by PVNLT \cite{Banerjee2016a, Banerjee2016b}. Finally, our principle can also be extended to study the turbulent relaxation of other non-trvial systems \textit{e.g.} compressible fluids and plasmas, ferrofluids and binary fluid systems where, unlike $\bu$ and $\bb$, the field variables are not necessarily solenoidal. \\

SB and AH contributed equally to this paper. 

SB aknowledges the support of CEFIPRA Project No. 6104-1 and also the DST INSPIRE faculty research grant (DST/PHY/2017514).


\begin{thebibliography}{42}%
\makeatletter
\providecommand \@ifxundefined [1]{%
 \@ifx{#1\undefined}
}%
\providecommand \@ifnum [1]{%
 \ifnum #1\expandafter \@firstoftwo
 \else \expandafter \@secondoftwo
 \fi
}%
\providecommand \@ifx [1]{%
 \ifx #1\expandafter \@firstoftwo
 \else \expandafter \@secondoftwo
 \fi
}%
\providecommand \natexlab [1]{#1}%
\providecommand \enquote  [1]{``#1''}%
\providecommand \bibnamefont  [1]{#1}%
\providecommand \bibfnamefont [1]{#1}%
\providecommand \citenamefont [1]{#1}%
\providecommand \href@noop [0]{\@secondoftwo}%
\providecommand \href [0]{\begingroup \@sanitize@url \@href}%
\providecommand \@href[1]{\@@startlink{#1}\@@href}%
\providecommand \@@href[1]{\endgroup#1\@@endlink}%
\providecommand \@sanitize@url [0]{\catcode `\\12\catcode `\$12\catcode
  `\&12\catcode `\#12\catcode `\^12\catcode `\_12\catcode `\%12\relax}%
\providecommand \@@startlink[1]{}%
\providecommand \@@endlink[0]{}%
\providecommand \url  [0]{\begingroup\@sanitize@url \@url }%
\providecommand \@url [1]{\endgroup\@href {#1}{\urlprefix }}%
\providecommand \urlprefix  [0]{URL }%
\providecommand \Eprint [0]{\href }%
\providecommand \doibase [0]{http://dx.doi.org/}%
\providecommand \selectlanguage [0]{\@gobble}%
\providecommand \bibinfo  [0]{\@secondoftwo}%
\providecommand \bibfield  [0]{\@secondoftwo}%
\providecommand \translation [1]{[#1]}%
\providecommand \BibitemOpen [0]{}%
\providecommand \bibitemStop [0]{}%
\providecommand \bibitemNoStop [0]{.\EOS\space}%
\providecommand \EOS [0]{\spacefactor3000\relax}%
\providecommand \BibitemShut  [1]{\csname bibitem#1\endcsname}%
\let\auto@bib@innerbib\@empty
\bibitem [{Note1()}]{Note1}%
  \BibitemOpen
  \bibinfo {note} {Expressed in Alfv\'en units}\BibitemShut {NoStop}%
\bibitem [{\citenamefont {Chandrasekhar}\ and\ \citenamefont
  {Woltjer}(1958)}]{Chandrasekhar1958}%
  \BibitemOpen
  \bibfield  {author} {\bibinfo {author} {\bibfnamefont {S.}~\bibnamefont
  {Chandrasekhar}}\ and\ \bibinfo {author} {\bibfnamefont {L.}~\bibnamefont
  {Woltjer}},\ }\href {\doibase 10.1073/pnas.44.4.285} {\bibfield  {journal}
  {\bibinfo  {journal} {Proceedings of the National Academy of Sciences}\
  }\textbf {\bibinfo {volume} {44}},\ \bibinfo {pages} {285} (\bibinfo {year}
  {1958})}\BibitemShut {NoStop}%
\bibitem [{\citenamefont {Woltjer}(1958{\natexlab{a}})}]{Woltjer1958a}%
  \BibitemOpen
  \bibfield  {author} {\bibinfo {author} {\bibfnamefont {L.}~\bibnamefont
  {Woltjer}},\ }\href {\doibase 10.1073/pnas.44.6.489} {\bibfield  {journal}
  {\bibinfo  {journal} {Proceedings of the National Academy of Sciences}\
  }\textbf {\bibinfo {volume} {44}},\ \bibinfo {pages} {489} (\bibinfo {year}
  {1958}{\natexlab{a}})}\BibitemShut {NoStop}%
\bibitem [{\citenamefont {Taylor}(1974)}]{Taylor1974}%
  \BibitemOpen
  \bibfield  {author} {\bibinfo {author} {\bibfnamefont {J.~B.}\ \bibnamefont
  {Taylor}},\ }\href {\doibase 10.1103/PhysRevLett.33.1139} {\bibfield
  {journal} {\bibinfo  {journal} {Phys. Rev. Lett.}\ }\textbf {\bibinfo
  {volume} {33}},\ \bibinfo {pages} {1139} (\bibinfo {year}
  {1974})}\BibitemShut {NoStop}%
\bibitem [{\citenamefont {Woltjer}(1958{\natexlab{b}})}]{Woltjer1958b}%
  \BibitemOpen
  \bibfield  {author} {\bibinfo {author} {\bibfnamefont {L.}~\bibnamefont
  {Woltjer}},\ }\href {\doibase 10.1073/pnas.44.9.833} {\bibfield  {journal}
  {\bibinfo  {journal} {Proceedings of the National Academy of Sciences}\
  }\textbf {\bibinfo {volume} {44}},\ \bibinfo {pages} {833} (\bibinfo {year}
  {1958}{\natexlab{b}})}\BibitemShut {NoStop}%
\bibitem [{\citenamefont {Mahajan}\ and\ \citenamefont
  {Yoshida}(1998)}]{Mahajan1998}%
  \BibitemOpen
  \bibfield  {author} {\bibinfo {author} {\bibfnamefont {S.~M.}\ \bibnamefont
  {Mahajan}}\ and\ \bibinfo {author} {\bibfnamefont {Z.}~\bibnamefont
  {Yoshida}},\ }\href {\doibase 10.1103/PhysRevLett.81.4863} {\bibfield
  {journal} {\bibinfo  {journal} {Phys. Rev. Lett.}\ }\textbf {\bibinfo
  {volume} {81}},\ \bibinfo {pages} {4863} (\bibinfo {year}
  {1998})}\BibitemShut {NoStop}%
\bibitem [{\citenamefont {Yoshida}\ and\ \citenamefont
  {Mahajan}(1999)}]{Yoshida1999}%
  \BibitemOpen
  \bibfield  {author} {\bibinfo {author} {\bibfnamefont {Z.}~\bibnamefont
  {Yoshida}}\ and\ \bibinfo {author} {\bibfnamefont {S.~M.}\ \bibnamefont
  {Mahajan}},\ }\href {\doibase 10.1063/1.533016} {\bibfield  {journal}
  {\bibinfo  {journal} {Journal of Mathematical Physics}\ }\textbf {\bibinfo
  {volume} {40}},\ \bibinfo {pages} {5080} (\bibinfo {year}
  {1999})}\BibitemShut {NoStop}%
\bibitem [{\citenamefont {Yoshida}\ and\ \citenamefont
  {Mahajan}(2002)}]{Yoshida2002}%
  \BibitemOpen
  \bibfield  {author} {\bibinfo {author} {\bibfnamefont {Z.}~\bibnamefont
  {Yoshida}}\ and\ \bibinfo {author} {\bibfnamefont {S.~M.}\ \bibnamefont
  {Mahajan}},\ }\href {\doibase 10.1103/PhysRevLett.88.095001} {\bibfield
  {journal} {\bibinfo  {journal} {Phys. Rev. Lett.}\ }\textbf {\bibinfo
  {volume} {88}},\ \bibinfo {pages} {095001} (\bibinfo {year}
  {2002})}\BibitemShut {NoStop}%
\bibitem [{\citenamefont {Qin}\ \emph {et~al.}(2012)\citenamefont {Qin},
  \citenamefont {Liu}, \citenamefont {Li},\ and\ \citenamefont
  {Squire}}]{Qin2012}%
  \BibitemOpen
  \bibfield  {author} {\bibinfo {author} {\bibfnamefont {H.}~\bibnamefont
  {Qin}}, \bibinfo {author} {\bibfnamefont {W.}~\bibnamefont {Liu}}, \bibinfo
  {author} {\bibfnamefont {H.}~\bibnamefont {Li}}, \ and\ \bibinfo {author}
  {\bibfnamefont {J.}~\bibnamefont {Squire}},\ }\href {\doibase
  10.1103/PhysRevLett.109.235001} {\bibfield  {journal} {\bibinfo  {journal}
  {Phys. Rev. Lett.}\ }\textbf {\bibinfo {volume} {109}},\ \bibinfo {pages}
  {235001} (\bibinfo {year} {2012})}\BibitemShut {NoStop}%
\bibitem [{\citenamefont {Zhu}\ \emph {et~al.}(1995)\citenamefont {Zhu},
  \citenamefont {Horiuchi}, \citenamefont {Sato},\ and\ \citenamefont
  {ComplexitySimulationGroup}}]{Zhu1995}%
  \BibitemOpen
  \bibfield  {author} {\bibinfo {author} {\bibfnamefont {S.~P.}\ \bibnamefont
  {Zhu}}, \bibinfo {author} {\bibfnamefont {R.}~\bibnamefont {Horiuchi}},
  \bibinfo {author} {\bibfnamefont {T.}~\bibnamefont {Sato}}, \ and\ \bibinfo
  {author} {\bibnamefont {ComplexitySimulationGroup}},\ }\href
  {https://link.aps.org/doi/10.1103/PhysRevE.51.6047} {\bibfield  {journal}
  {\bibinfo  {journal} {Phys. Rev. E}\ }\textbf {\bibinfo {volume} {51}},\
  \bibinfo {pages} {6047} (\bibinfo {year} {1995})}\BibitemShut {NoStop}%
\bibitem [{\citenamefont {Zhu}\ \emph {et~al.}(1996)\citenamefont {Zhu},
  \citenamefont {Horiuchi},\ and\ \citenamefont {Sato}}]{Zhu1996}%
  \BibitemOpen
  \bibfield  {author} {\bibinfo {author} {\bibfnamefont {S.~P.}\ \bibnamefont
  {Zhu}}, \bibinfo {author} {\bibfnamefont {R.}~\bibnamefont {Horiuchi}}, \
  and\ \bibinfo {author} {\bibfnamefont {T.}~\bibnamefont {Sato}},\ }\href
  {\doibase 10.1063/1.871717} {\bibfield  {journal} {\bibinfo  {journal}
  {Physics of Plasmas}\ }\textbf {\bibinfo {volume} {3}},\ \bibinfo {pages}
  {2821} (\bibinfo {year} {1996})}\BibitemShut {NoStop}%
\bibitem [{\citenamefont {Sato}(1996)}]{Sato1996}%
  \BibitemOpen
  \bibfield  {author} {\bibinfo {author} {\bibfnamefont {T.}~\bibnamefont
  {Sato}},\ }\href {\doibase 10.1063/1.871666} {\bibfield  {journal} {\bibinfo
  {journal} {Physics of Plasmas}\ }\textbf {\bibinfo {volume} {3}},\ \bibinfo
  {pages} {2135} (\bibinfo {year} {1996})}\BibitemShut {NoStop}%
\bibitem [{\citenamefont {Prigogine}(1955)}]{Prigogine1955}%
  \BibitemOpen
  \bibfield  {author} {\bibinfo {author} {\bibfnamefont {I.}~\bibnamefont
  {Prigogine}},\ }\href {https://books.google.co.in/books?id=wAbwAAAAMAAJ}
  {\emph {\bibinfo {title} {Introduction to Thermodynamics of Irreversible
  Processes}}},\ American lecture series\ (\bibinfo  {publisher} {Thomas},\
  \bibinfo {year} {1955})\BibitemShut {NoStop}%
\bibitem [{\citenamefont {Ting}\ \emph {et~al.}(1986)\citenamefont {Ting},
  \citenamefont {Matthaeus},\ and\ \citenamefont {Montgomery}}]{Ting1986}%
  \BibitemOpen
  \bibfield  {author} {\bibinfo {author} {\bibfnamefont {A.~C.}\ \bibnamefont
  {Ting}}, \bibinfo {author} {\bibfnamefont {W.~H.}\ \bibnamefont {Matthaeus}},
  \ and\ \bibinfo {author} {\bibfnamefont {D.}~\bibnamefont {Montgomery}},\
  }\href {\doibase 10.1063/1.865843} {\bibfield  {journal} {\bibinfo  {journal}
  {The Physics of Fluids}\ }\textbf {\bibinfo {volume} {29}},\ \bibinfo {pages}
  {3261} (\bibinfo {year} {1986})}\BibitemShut {NoStop}%
\bibitem [{\citenamefont {Hameiri}\ and\ \citenamefont
  {Bhattacharjee}(1987)}]{Hamieri1987}%
  \BibitemOpen
  \bibfield  {author} {\bibinfo {author} {\bibfnamefont {E.}~\bibnamefont
  {Hameiri}}\ and\ \bibinfo {author} {\bibfnamefont {A.}~\bibnamefont
  {Bhattacharjee}},\ }\href {\doibase 10.1103/PhysRevA.35.768} {\bibfield
  {journal} {\bibinfo  {journal} {Phys. Rev. A}\ }\textbf {\bibinfo {volume}
  {35}},\ \bibinfo {pages} {768} (\bibinfo {year} {1987})}\BibitemShut
  {NoStop}%
\bibitem [{\citenamefont {Montgomery}\ and\ \citenamefont
  {Phillips}(1988)}]{Montgomery1988}%
  \BibitemOpen
  \bibfield  {author} {\bibinfo {author} {\bibfnamefont {D.}~\bibnamefont
  {Montgomery}}\ and\ \bibinfo {author} {\bibfnamefont {L.}~\bibnamefont
  {Phillips}},\ }\href {\doibase 10.1103/PhysRevA.38.2953} {\bibfield
  {journal} {\bibinfo  {journal} {Phys. Rev. A}\ }\textbf {\bibinfo {volume}
  {38}},\ \bibinfo {pages} {2953} (\bibinfo {year} {1988})}\BibitemShut
  {NoStop}%
\bibitem [{\citenamefont {Montgomery}\ and\ \citenamefont
  {Phillips}(1989)}]{Montgomery1989}%
  \BibitemOpen
  \bibfield  {author} {\bibinfo {author} {\bibfnamefont {D.}~\bibnamefont
  {Montgomery}}\ and\ \bibinfo {author} {\bibfnamefont {L.}~\bibnamefont
  {Phillips}},\ }\href {\doibase https://doi.org/10.1016/0167-2789(89)90131-0}
  {\bibfield  {journal} {\bibinfo  {journal} {Physica D: Nonlinear Phenomena}\
  }\textbf {\bibinfo {volume} {37}},\ \bibinfo {pages} {215} (\bibinfo {year}
  {1989})}\BibitemShut {NoStop}%
\bibitem [{\citenamefont {Dasgupta}\ \emph {et~al.}(1998)\citenamefont
  {Dasgupta}, \citenamefont {Dasgupta}, \citenamefont {Janaki}, \citenamefont
  {Watanabe},\ and\ \citenamefont {Sato}}]{Dasgupta1998}%
  \BibitemOpen
  \bibfield  {author} {\bibinfo {author} {\bibfnamefont {B.}~\bibnamefont
  {Dasgupta}}, \bibinfo {author} {\bibfnamefont {P.}~\bibnamefont {Dasgupta}},
  \bibinfo {author} {\bibfnamefont {M.~S.}\ \bibnamefont {Janaki}}, \bibinfo
  {author} {\bibfnamefont {T.}~\bibnamefont {Watanabe}}, \ and\ \bibinfo
  {author} {\bibfnamefont {T.}~\bibnamefont {Sato}},\ }\href {\doibase
  10.1103/PhysRevLett.81.3144} {\bibfield  {journal} {\bibinfo  {journal}
  {Phys. Rev. Lett.}\ }\textbf {\bibinfo {volume} {81}},\ \bibinfo {pages}
  {3144} (\bibinfo {year} {1998})}\BibitemShut {NoStop}%
\bibitem [{\citenamefont {González}\ \emph {et~al.}(2008)\citenamefont
  {González}, \citenamefont {Sarasua},\ and\ \citenamefont
  {Costa}}]{Gonzalez2008}%
  \BibitemOpen
  \bibfield  {author} {\bibinfo {author} {\bibfnamefont {R.}~\bibnamefont
  {González}}, \bibinfo {author} {\bibfnamefont {G.}~\bibnamefont {Sarasua}},
  \ and\ \bibinfo {author} {\bibfnamefont {A.}~\bibnamefont {Costa}},\ }\href
  {\doibase 10.1063/1.2840196} {\bibfield  {journal} {\bibinfo  {journal}
  {Physics of Fluids}\ }\textbf {\bibinfo {volume} {20}},\ \bibinfo {pages}
  {024106} (\bibinfo {year} {2008})}\BibitemShut {NoStop}%
\bibitem [{\citenamefont {González}\ \emph {et~al.}(2010)\citenamefont
  {González}, \citenamefont {Costa},\ and\ \citenamefont
  {Santini}}]{Gonzalez2010}%
  \BibitemOpen
  \bibfield  {author} {\bibinfo {author} {\bibfnamefont {R.}~\bibnamefont
  {González}}, \bibinfo {author} {\bibfnamefont {A.}~\bibnamefont {Costa}}, \
  and\ \bibinfo {author} {\bibfnamefont {E.~S.}\ \bibnamefont {Santini}},\
  }\href {\doibase 10.1063/1.3460297} {\bibfield  {journal} {\bibinfo
  {journal} {Physics of Fluids}\ }\textbf {\bibinfo {volume} {22}},\ \bibinfo
  {pages} {074102} (\bibinfo {year} {2010})}\BibitemShut {NoStop}%
\bibitem [{\citenamefont {Kraichnan}\ and\ \citenamefont
  {Panda}(1988)}]{Kraichnan1988}%
  \BibitemOpen
  \bibfield  {author} {\bibinfo {author} {\bibfnamefont {R.~H.}\ \bibnamefont
  {Kraichnan}}\ and\ \bibinfo {author} {\bibfnamefont {R.}~\bibnamefont
  {Panda}},\ }\href {\doibase 10.1063/1.866591} {\bibfield  {journal} {\bibinfo
   {journal} {The Physics of Fluids}\ }\textbf {\bibinfo {volume} {31}},\
  \bibinfo {pages} {2395} (\bibinfo {year} {1988})}\BibitemShut {NoStop}%
\bibitem [{\citenamefont {She}\ \emph {et~al.}(1991)\citenamefont {She},
  \citenamefont {Jackson},\ and\ \citenamefont {Orszag}}]{She1991}%
  \BibitemOpen
  \bibfield  {author} {\bibinfo {author} {\bibfnamefont {Z.-S.}\ \bibnamefont
  {She}}, \bibinfo {author} {\bibfnamefont {E.}~\bibnamefont {Jackson}}, \ and\
  \bibinfo {author} {\bibfnamefont {S.~A.}\ \bibnamefont {Orszag}},\ }\href
  {\doibase 10.1098/rspa.1991.0083} {\bibfield  {journal} {\bibinfo  {journal}
  {Proceedings of the Royal Society of London. Series A: Mathematical and
  Physical Sciences}\ }\textbf {\bibinfo {volume} {434}},\ \bibinfo {pages}
  {101} (\bibinfo {year} {1991})}\BibitemShut {NoStop}%
\bibitem [{\citenamefont {Matthaeus}\ and\ \citenamefont
  {Montgomery}(1980)}]{Matthaeus1980}%
  \BibitemOpen
  \bibfield  {author} {\bibinfo {author} {\bibfnamefont {W.~H.}\ \bibnamefont
  {Matthaeus}}\ and\ \bibinfo {author} {\bibfnamefont {D.}~\bibnamefont
  {Montgomery}},\ }\href {\doibase
  https://doi.org/10.1111/j.1749-6632.1980.tb29687.x} {\bibfield  {journal}
  {\bibinfo  {journal} {Annals of the New York Academy of Sciences}\ }\textbf
  {\bibinfo {volume} {357}},\ \bibinfo {pages} {203} (\bibinfo {year}
  {1980})}\BibitemShut {NoStop}%
\bibitem [{\citenamefont {Stribling}\ and\ \citenamefont
  {Matthaeus}(1991)}]{Stribling1991}%
  \BibitemOpen
  \bibfield  {author} {\bibinfo {author} {\bibfnamefont {T.}~\bibnamefont
  {Stribling}}\ and\ \bibinfo {author} {\bibfnamefont {W.~H.}\ \bibnamefont
  {Matthaeus}},\ }\href {\doibase 10.1063/1.859654} {\bibfield  {journal}
  {\bibinfo  {journal} {Physics of Fluids B: Plasma Physics}\ }\textbf
  {\bibinfo {volume} {3}},\ \bibinfo {pages} {1848} (\bibinfo {year}
  {1991})}\BibitemShut {NoStop}%
\bibitem [{\citenamefont {Servidio}\ \emph {et~al.}(2008)\citenamefont
  {Servidio}, \citenamefont {Matthaeus},\ and\ \citenamefont
  {Dmitruk}}]{Servidio2008}%
  \BibitemOpen
  \bibfield  {author} {\bibinfo {author} {\bibfnamefont {S.}~\bibnamefont
  {Servidio}}, \bibinfo {author} {\bibfnamefont {W.~H.}\ \bibnamefont
  {Matthaeus}}, \ and\ \bibinfo {author} {\bibfnamefont {P.}~\bibnamefont
  {Dmitruk}},\ }\href {\doibase 10.1103/PhysRevLett.100.095005} {\bibfield
  {journal} {\bibinfo  {journal} {Phys. Rev. Lett.}\ }\textbf {\bibinfo
  {volume} {100}},\ \bibinfo {pages} {095005} (\bibinfo {year}
  {2008})}\BibitemShut {NoStop}%
\bibitem [{\citenamefont {Tsinober}\ \emph {et~al.}(1999)\citenamefont
  {Tsinober}, \citenamefont {Ortenberg},\ and\ \citenamefont
  {Shtilman}}]{Tsinober1999}%
  \BibitemOpen
  \bibfield  {author} {\bibinfo {author} {\bibfnamefont {A.}~\bibnamefont
  {Tsinober}}, \bibinfo {author} {\bibfnamefont {M.}~\bibnamefont {Ortenberg}},
  \ and\ \bibinfo {author} {\bibfnamefont {L.}~\bibnamefont {Shtilman}},\
  }\href {\doibase 10.1063/1.870091} {\bibfield  {journal} {\bibinfo  {journal}
  {Physics of Fluids}\ }\textbf {\bibinfo {volume} {11}},\ \bibinfo {pages}
  {2291} (\bibinfo {year} {1999})}\BibitemShut {NoStop}%
\bibitem [{\citenamefont {Monin}\ and\ \citenamefont
  {Yaglom}(1975)}]{Monin1975}%
  \BibitemOpen
  \bibfield  {author} {\bibinfo {author} {\bibfnamefont {A.~S.}\ \bibnamefont
  {Monin}}\ and\ \bibinfo {author} {\bibfnamefont {A.~M.}\ \bibnamefont
  {Yaglom}},\ }\href@noop {} {\emph {\bibinfo {title} {Statistical Fluid
  Mechanics: Mechanics of Turbulence}}}\ (\bibinfo  {publisher} {Mit Press},\
  \bibinfo {year} {1975})\BibitemShut {NoStop}%
\bibitem [{\citenamefont {Politano}\ and\ \citenamefont
  {Pouquet}(1998)}]{Politano1998b}%
  \BibitemOpen
  \bibfield  {author} {\bibinfo {author} {\bibfnamefont {H.}~\bibnamefont
  {Politano}}\ and\ \bibinfo {author} {\bibfnamefont {A.}~\bibnamefont
  {Pouquet}},\ }\href {\doibase https://doi.org/10.1029/97GL03642} {\bibfield
  {journal} {\bibinfo  {journal} {Geophysical Research Letters}\ }\textbf
  {\bibinfo {volume} {25}},\ \bibinfo {pages} {273} (\bibinfo {year}
  {1998})}\BibitemShut {NoStop}%
\bibitem [{\citenamefont {Banerjee}\ and\ \citenamefont
  {Galtier}(2016{\natexlab{a}})}]{Banerjee2016a}%
  \BibitemOpen
  \bibfield  {author} {\bibinfo {author} {\bibfnamefont {S.}~\bibnamefont
  {Banerjee}}\ and\ \bibinfo {author} {\bibfnamefont {S.}~\bibnamefont
  {Galtier}},\ }\href {\doibase 10.1088/1751-8113/50/1/015501} {\bibfield
  {journal} {\bibinfo  {journal} {J. Phys. A: Math. Theor.}\ }\textbf {\bibinfo
  {volume} {50}},\ \bibinfo {pages} {015501} (\bibinfo {year}
  {2016}{\natexlab{a}})}\BibitemShut {NoStop}%
\bibitem [{\citenamefont {Banerjee}\ and\ \citenamefont
  {Galtier}(2016{\natexlab{b}})}]{Banerjee2016b}%
  \BibitemOpen
  \bibfield  {author} {\bibinfo {author} {\bibfnamefont {S.}~\bibnamefont
  {Banerjee}}\ and\ \bibinfo {author} {\bibfnamefont {S.}~\bibnamefont
  {Galtier}},\ }\href {\doibase 10.1103/PhysRevE.93.033120} {\bibfield
  {journal} {\bibinfo  {journal} {Phys. Rev. E}\ }\textbf {\bibinfo {volume}
  {93}},\ \bibinfo {pages} {033120} (\bibinfo {year}
  {2016}{\natexlab{b}})}\BibitemShut {NoStop}%
\bibitem [{\citenamefont {Mouraya}\ and\ \citenamefont
  {Banerjee}(2019)}]{Mouraya2019}%
  \BibitemOpen
  \bibfield  {author} {\bibinfo {author} {\bibfnamefont {S.}~\bibnamefont
  {Mouraya}}\ and\ \bibinfo {author} {\bibfnamefont {S.}~\bibnamefont
  {Banerjee}},\ }\href {\doibase 10.1103/PhysRevE.100.053105} {\bibfield
  {journal} {\bibinfo  {journal} {Phys. Rev. E}\ }\textbf {\bibinfo {volume}
  {100}},\ \bibinfo {pages} {053105} (\bibinfo {year} {2019})}\BibitemShut
  {NoStop}%
\bibitem [{\citenamefont {Pan}\ and\ \citenamefont {Banerjee}(2022)}]{Pan2022}%
  \BibitemOpen
  \bibfield  {author} {\bibinfo {author} {\bibfnamefont {N.}~\bibnamefont
  {Pan}}\ and\ \bibinfo {author} {\bibfnamefont {S.}~\bibnamefont {Banerjee}},\
  }\href {\doibase 10.1103/PhysRevE.106.025104} {\bibfield  {journal} {\bibinfo
   {journal} {Phys. Rev. E}\ }\textbf {\bibinfo {volume} {106}},\ \bibinfo
  {pages} {025104} (\bibinfo {year} {2022})}\BibitemShut {NoStop}%
\bibitem [{Note2()}]{Note2}%
  \BibitemOpen
  \bibinfo {note} {The Imposition of a cutoff is consistent with the
  dissipative anomaly of turbulence theory.}\BibitemShut {Stop}%
\bibitem [{\citenamefont {Carnevale}\ \emph {et~al.}(1981)\citenamefont
  {Carnevale}, \citenamefont {Frisch},\ and\ \citenamefont
  {Salmon}}]{Carnevale1981}%
  \BibitemOpen
  \bibfield  {author} {\bibinfo {author} {\bibfnamefont {G.~F.}\ \bibnamefont
  {Carnevale}}, \bibinfo {author} {\bibfnamefont {U.}~\bibnamefont {Frisch}}, \
  and\ \bibinfo {author} {\bibfnamefont {R.}~\bibnamefont {Salmon}},\ }\href
  {\doibase 10.1088/0305-4470/14/7/026} {\bibfield  {journal} {\bibinfo
  {journal} {Journal of Physics A: Mathematical and General}\ }\textbf
  {\bibinfo {volume} {14}},\ \bibinfo {pages} {1701} (\bibinfo {year}
  {1981})}\BibitemShut {NoStop}%
\bibitem [{\citenamefont {Riley}\ \emph {et~al.}(1995)\citenamefont {Riley},
  \citenamefont {Sonett}, \citenamefont {Balogh}, \citenamefont {Forsyth},
  \citenamefont {Scime},\ and\ \citenamefont {Feldman}}]{Riley1995}%
  \BibitemOpen
  \bibfield  {author} {\bibinfo {author} {\bibfnamefont {P.}~\bibnamefont
  {Riley}}, \bibinfo {author} {\bibfnamefont {C.}~\bibnamefont {Sonett}},
  \bibinfo {author} {\bibfnamefont {A.}~\bibnamefont {Balogh}}, \bibinfo
  {author} {\bibfnamefont {R.}~\bibnamefont {Forsyth}}, \bibinfo {author}
  {\bibfnamefont {E.}~\bibnamefont {Scime}}, \ and\ \bibinfo {author}
  {\bibfnamefont {W.}~\bibnamefont {Feldman}},\ }\href@noop {} {\bibfield
  {journal} {\bibinfo  {journal} {Space Science Reviews}\ }\textbf {\bibinfo
  {volume} {72}},\ \bibinfo {pages} {197} (\bibinfo {year} {1995})}\BibitemShut
  {NoStop}%
\bibitem [{\citenamefont {Wicks}\ \emph {et~al.}(2013)\citenamefont {Wicks},
  \citenamefont {Mallet}, \citenamefont {Horbury}, \citenamefont {Chen},
  \citenamefont {Schekochihin},\ and\ \citenamefont {Mitchell}}]{Wicks2013}%
  \BibitemOpen
  \bibfield  {author} {\bibinfo {author} {\bibfnamefont {R.~T.}\ \bibnamefont
  {Wicks}}, \bibinfo {author} {\bibfnamefont {A.}~\bibnamefont {Mallet}},
  \bibinfo {author} {\bibfnamefont {T.~S.}\ \bibnamefont {Horbury}}, \bibinfo
  {author} {\bibfnamefont {C.~H.~K.}\ \bibnamefont {Chen}}, \bibinfo {author}
  {\bibfnamefont {A.~A.}\ \bibnamefont {Schekochihin}}, \ and\ \bibinfo
  {author} {\bibfnamefont {J.~J.}\ \bibnamefont {Mitchell}},\ }\href {\doibase
  10.1103/PhysRevLett.110.025003} {\bibfield  {journal} {\bibinfo  {journal}
  {Phys. Rev. Lett.}\ }\textbf {\bibinfo {volume} {110}},\ \bibinfo {pages}
  {025003} (\bibinfo {year} {2013})}\BibitemShut {NoStop}%
\bibitem [{\citenamefont {Bhattacharyya}\ \emph {et~al.}(2003)\citenamefont
  {Bhattacharyya}, \citenamefont {Janaki},\ and\ \citenamefont
  {Dasgupta}}]{Bhattacharyya2003}%
  \BibitemOpen
  \bibfield  {author} {\bibinfo {author} {\bibfnamefont {R.}~\bibnamefont
  {Bhattacharyya}}, \bibinfo {author} {\bibfnamefont {M.}~\bibnamefont
  {Janaki}}, \ and\ \bibinfo {author} {\bibfnamefont {B.}~\bibnamefont
  {Dasgupta}},\ }\href {\doibase 10.1016/S0375-9601(03)00977-0} {\bibfield
  {journal} {\bibinfo  {journal} {Physics Letters A}\ }\textbf {\bibinfo
  {volume} {315}},\ \bibinfo {pages} {120} (\bibinfo {year}
  {2003})}\BibitemShut {NoStop}%
\bibitem [{\citenamefont {Hasegawa}(1985)}]{Hasegawa1985}%
  \BibitemOpen
  \bibfield  {author} {\bibinfo {author} {\bibfnamefont {A.}~\bibnamefont
  {Hasegawa}},\ }\href {\doibase 10.1080/00018738500101721} {\bibfield
  {journal} {\bibinfo  {journal} {Advances in Physics}\ }\textbf {\bibinfo
  {volume} {34}},\ \bibinfo {pages} {1} (\bibinfo {year} {1985})}\BibitemShut
  {NoStop}%
\bibitem [{Note3()}]{Note3}%
  \BibitemOpen
  \bibinfo {note} {Following \cite {Yoshida2002}, a richer class of relaxed
  states may be obtained if the palinstrophy $P\ (=\DOTSI \intop \ilimits@
  \left (\nabla \times {\protect \bm {\omega }} \right )^2 d \tau )$ is varied
  for a fixed value of $E_K$ and $\Omega $. Such a variation would lead to a
  quadruple-curl Beltrami state in ${\protect \bm {u}}$ given by $\nabla \times
  \nabla \times \left ( \nabla \times \protect \bm {\omega }\right )= {\lambda
  _1} {{\protect \bm {u}}} + {\lambda _2} \left (\nabla \times \protect \bm
  {\omega }\right )$ which permits Eq.~\protect \eqref {align2d1} as a possible
  solution.}\BibitemShut {Stop}%
\bibitem [{\citenamefont {Donato}\ \emph {et~al.}(2012)\citenamefont {Donato},
  \citenamefont {Servidio}, \citenamefont {Dmitruk}, \citenamefont {Carbone},
  \citenamefont {Shay}, \citenamefont {Cassak},\ and\ \citenamefont
  {Matthaeus}}]{Donato2012}%
  \BibitemOpen
  \bibfield  {author} {\bibinfo {author} {\bibfnamefont {S.}~\bibnamefont
  {Donato}}, \bibinfo {author} {\bibfnamefont {S.}~\bibnamefont {Servidio}},
  \bibinfo {author} {\bibfnamefont {P.}~\bibnamefont {Dmitruk}}, \bibinfo
  {author} {\bibfnamefont {V.}~\bibnamefont {Carbone}}, \bibinfo {author}
  {\bibfnamefont {M.~A.}\ \bibnamefont {Shay}}, \bibinfo {author}
  {\bibfnamefont {P.~A.}\ \bibnamefont {Cassak}}, \ and\ \bibinfo {author}
  {\bibfnamefont {W.~H.}\ \bibnamefont {Matthaeus}},\ }\href {\doibase
  10.1063/1.4754151} {\bibfield  {journal} {\bibinfo  {journal} {Physics of
  Plasmas}\ }\textbf {\bibinfo {volume} {19}},\ \bibinfo {pages} {092307}
  (\bibinfo {year} {2012})}\BibitemShut {NoStop}%
\bibitem [{\citenamefont {Wang}\ \emph {et~al.}(2012)\citenamefont {Wang},
  \citenamefont {Xiao},\ and\ \citenamefont {Wang}}]{Wang2012}%
  \BibitemOpen
  \bibfield  {author} {\bibinfo {author} {\bibfnamefont {J.}~\bibnamefont
  {Wang}}, \bibinfo {author} {\bibfnamefont {C.}~\bibnamefont {Xiao}}, \ and\
  \bibinfo {author} {\bibfnamefont {X.}~\bibnamefont {Wang}},\ }\href {\doibase
  10.1063/1.3697561} {\bibfield  {journal} {\bibinfo  {journal} {Physics of
  Plasmas}\ }\textbf {\bibinfo {volume} {19}},\ \bibinfo {pages} {032905}
  (\bibinfo {year} {2012})}\BibitemShut {NoStop}%
\bibitem [{\citenamefont {Papini}\ \emph {et~al.}(2021)\citenamefont {Papini},
  \citenamefont {Hellinger}, \citenamefont {Verdini}, \citenamefont {Landi},
  \citenamefont {Franci}, \citenamefont {Montagud-Camps},\ and\ \citenamefont
  {Matteini}}]{Papini2021}%
  \BibitemOpen
  \bibfield  {author} {\bibinfo {author} {\bibfnamefont {E.}~\bibnamefont
  {Papini}}, \bibinfo {author} {\bibfnamefont {P.}~\bibnamefont {Hellinger}},
  \bibinfo {author} {\bibfnamefont {A.}~\bibnamefont {Verdini}}, \bibinfo
  {author} {\bibfnamefont {S.}~\bibnamefont {Landi}}, \bibinfo {author}
  {\bibfnamefont {L.}~\bibnamefont {Franci}}, \bibinfo {author} {\bibfnamefont
  {V.}~\bibnamefont {Montagud-Camps}}, \ and\ \bibinfo {author} {\bibfnamefont
  {L.}~\bibnamefont {Matteini}},\ }\href {\doibase 10.3390/atmos12121632}
  {\bibfield  {journal} {\bibinfo  {journal} {Atmosphere}\ }\textbf {\bibinfo
  {volume} {12}} (\bibinfo {year} {2021}),\ 10.3390/atmos12121632}\BibitemShut
  {NoStop}%
\end{thebibliography}
\end{document}